%% file: acl_latex.tex
\pdfoutput=1

\documentclass[11pt]{article}

\usepackage[final]{acl}

\usepackage{times}
\usepackage{latexsym}

\usepackage[T1]{fontenc}

\usepackage[utf8]{inputenc}

\usepackage{microtype}

\usepackage{inconsolata}

\usepackage{graphicx}

\usepackage{amsthm,amsmath,amssymb}
\usepackage{mathrsfs}

\usepackage{algorithm}
\usepackage{algorithmicx}
\usepackage[noend]{algpseudocode}

\usepackage{multirow}
\usepackage{microtype}      
\usepackage{booktabs}
\usepackage{adjustbox}

\usepackage{graphicx}
\usepackage{slashed}

\usepackage{makecell}

\title{DiffZOO: A Purely Query-Based Black-Box Attack for Red-teaming Text-to-Image Generative Model via Zeroth Order Optimization\\
\vspace{3mm}
\small{\textcolor{red}{\textmd{WARNING: This paper contains model-generated content that may be offensive or upsetting.}}}
}

\author{
  \textbf{\small{Pucheng Dang}} \textsuperscript{1,2}, 
  \textbf{\small{Xing Hu}} \textsuperscript{1,4}, 
  \textbf{\small{Dong Li \textsuperscript{1}}}, 
  \textbf{\small{Rui Zhang \textsuperscript{1}}}, 
  \textbf{\small{Qi Guo \textsuperscript{1}}},
  \textbf{\small{Kaidi Xu \textsuperscript{3}}}\\
\small{\textsuperscript{1} State Key Lab of Processors, Institute of Computing Technology, Chinese Academy of Sciences, Beijing, China} \\
\small{\textsuperscript{2} University of Chinese Academy of Sciences, China}, 
\small{\textsuperscript{3} Drexel University} \\
\small{\textsuperscript{4} Shanghai Innovation Center for Processor Technologies, SHIC} \\
\texttt{\small{\{dangpucheng20g,huxing,lidong,zhangrui,guoqi\}@ict.ac.cn} }\\
\texttt{\small{\{kx46\}@drexel.edu}
}}

\begin{document}
\maketitle
\input{Sections/abstract}

\input{Sections/introduction}

\input{Sections/related_works}

\input{Sections/preliminary}

\input{Sections/method}

\input{Sections/experiments}

\input{Sections/conclusion}


\input{Sections/limitations}

\bibliography{custom}

\clearpage
\appendix

\input{Sections/appendix}

\end{document}

%% file: Sections/abstract.tex
\begin{abstract}
Current text-to-image (T2I) synthesis diffusion models raise misuse concerns, particularly in creating prohibited or not-safe-for-work (NSFW) images. To address this, various safety mechanisms and red teaming attack methods are proposed to enhance or expose the T2I model's capability to generate unsuitable content. However, many red teaming attack methods assume knowledge of the text encoders, limiting their practical usage. In this work, we rethink the case of \textit{purely black-box} attacks without prior knowledge of the T2l model. 
To overcome the unavailability of gradients and the inability to optimize attacks within a discrete prompt space, we propose \textit{DiffZOO} which applies Zeroth Order Optimization to procure gradient approximations and harnesses both C-PRV and D-PRV to enhance attack prompts within the discrete prompt domain.
We evaluated our method across multiple safety mechanisms of the T2I diffusion model and online servers. Experiments on multiple state-of-the-art safety mechanisms show that \textit{DiffZOO} attains an 8.5\% higher average attack success rate than previous works, hence its promise as a practical red teaming tool for T2l models. Our code is available at \url{https://github.com/CherryBlueberry/DiffZOO}.
\end{abstract}

%% file: Sections/introduction.tex
\section{Introduction}
The domain of Generative AI has witnessed remarkable strides, notably in the realms of text \cite{Intro-1}, image \cite{Intro-2}, and code synthesis \cite{Intro-3}. Prominently, text-to-image (T2I) generation has risen as a central focus of investigations. The triumph of today's T2I diffusion models is significantly underpinned by the expansive online datasets utilized for their training. While this wealth of data enables T2I models to conjure a wide array of realistic imagery, it concurrently introduces challenges. Predominantly, the presence of sensitive content in images sourced from the web can result in trained models unintentionally internalizing and regurgitating unsuitable visuals. This encompasses concerns such as copyright breaches \cite{copyright}, images harboring forbidden content \cite{unlearning}, and NSFW (Not Safe For Work) materials \cite{sld}.

\begin{figure}[t]
\begin{center}
\centerline{\includegraphics[width=\columnwidth,height=0.574\columnwidth]{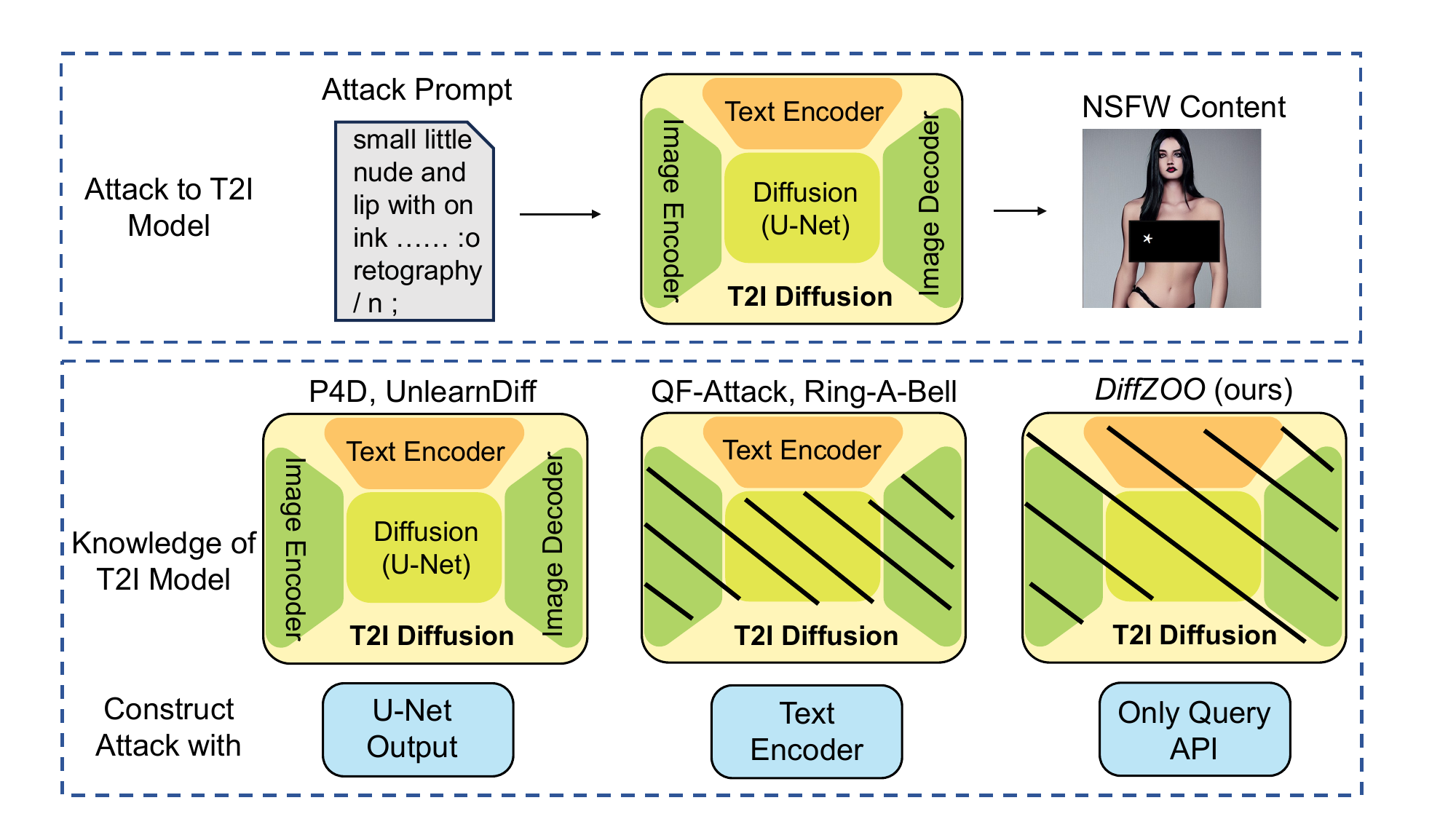}}
\vskip -0.1in
\caption{\footnotesize{\textit{First row}: The pipeline of attack methods that aim to evaluate the T2I diffusion model's safety mechanisms to find problematic prompts with the ability to reveal inappropriate concepts (such as “nudity” and “violence”). \textit{Second row}: our black-box attack \textit{DiffZOO} discards the shortcomings of the previous work and constructs attack prompts by purely querying the T2I generative model's API. 
}}
\label{show}
\end{center}
\vskip -0.5in
\end{figure}

To mitigate the generation of images containing inappropriate content, various safety mechanisms have been incorporated into diffusion models \cite{sld, esd, ca, fmn}. To evaluate these safety mechanisms, recent research proposes attack methods \cite{qf-attack,p4d,ring-a-bell,jap-Attack,unlearndiff} that aim to facilitate the red-teaming of T2I diffusion models equipped with safety mechanisms to find problematic prompts with the ability to reveal inappropriate concepts (e.g., craft attack prompts capable of generating images with banned concepts such as “nudity” and “violence”). 

These attack methods can be bifurcated into two primary categories: white-box attacks \cite{p4d,unlearndiff} and black-box attacks \cite{qf-attack,jap-Attack,ring-a-bell}. 
In the domain of white-box attacks, it is presupposed that the attacker has cognizance of the diffusion model's U-Net \cite{dm} output. Nonetheless, in genuine attack situations, the attacker cannot obtain the noise estimation of the U-Net output, which is essential for devising attack prompts.
Conversely, black-box attacks contrive attack prompts by capitalizing on Contrastive Language-Image Pre-training (CLIP) \cite{clip} embedding vectors. 
Nonetheless, this approach does not fully align with the orthodox definition of a black-box attack, considering the CLIP encoder's intrinsic role within the T2I diffusion model. 
Simultaneously, this attack paradigm shifts into a transferable attack when the targeted diffusion model resorts to a non-CLIP text encoder for generating embedding vectors.

In this work, we rethink the black-box attack for the T2I diffusion model to expose their propensity to generate inappropriate concepts. The attacker is restricted to querying the diffusion model through prompts and harnessing the generated images to develop attack prompts \cite{query-attack1, query-attack2}. In this case, we propose a query-based attack method, \textit{DiffZOO}, specifically tailored for the T2I diffusion model. There are two significant challenges in accomplishing this task. The \textbf{first} challenge is the inability to acquire gradients to construct the attack prompts, as is typically done in conventional adversarial attacks. To address this issue, we employ Zeroth Order Optimization (ZOO) \cite{zo-adamm, zoo-w, deep-zero} and obtain gradient estimates to construct attack prompts. The \textbf{second} challenge is distinguishing our work from previous studies of black-box attack \cite{qf-attack,ring-a-bell,jap-Attack}, where we consider the text encoder (CLIP) in a black-box setting (as shown in Figure \ref{show}). This scenario presents a disparity between the continuous embedding domain and the discrete prompt (token) domain. The former is naturally suited for optimizing an attack vector, as in conventional adversarial attack \cite{PGD,C&W,auto-attack}. However, optimizing the attack prompt on a discrete prompt domain presents a formidable problem. Fortunately, taking inspiration from TextGrad \cite{textgrad}, we utilize continuous position replacement vectors (C-PRV) to overcome such a problem. We subsequently sample from it to derive discrete position replacement vectors (D-PRV) and construct attack prompts without a text encoder. 

Our approach discerns the necessity for token replacement within prompts and selects suitable synonyms for crafting attack prompts. This approach successfully bridges the gap and facilitates direct optimization of the discrete prompt (token) domain. We summarize our contributions below. 
\begin{itemize}
\item We rethink the black-box settings of the T2I diffusion model, and regard the text encoder and the whole model as a black-box setting (previous black-box attacks regard the text encoder as a white-box setting), which requires no prior knowledge of the T2l model.
\item To overcome the unavailability of gradients and the inability to optimize attacks within a discrete prompt domain, we propose \textit{DiffZOO} a query-based attack method that serves as a prompt-based concept testing framework for red-team T2I diffusion models. \textit{DiffZOO} applies Zeroth Order Optimization to procure gradient approximations and harnesses both C-PRV and D-PRV to enhance attack prompts within the discrete prompt domain.
\item Our comprehensive experiments evaluate a wide range of models, encompassing prevalent online services to state-of-the-art methods in concept removal. Results reveal that prompts crafted by \textit{DiffZOO} significantly boost the average success rate of concept removal methods producing inappropriate images, an elevation beyond 8.5\%.
\end{itemize}
\begin{figure*}[t]
\begin{center}
	\includegraphics[width=\linewidth]{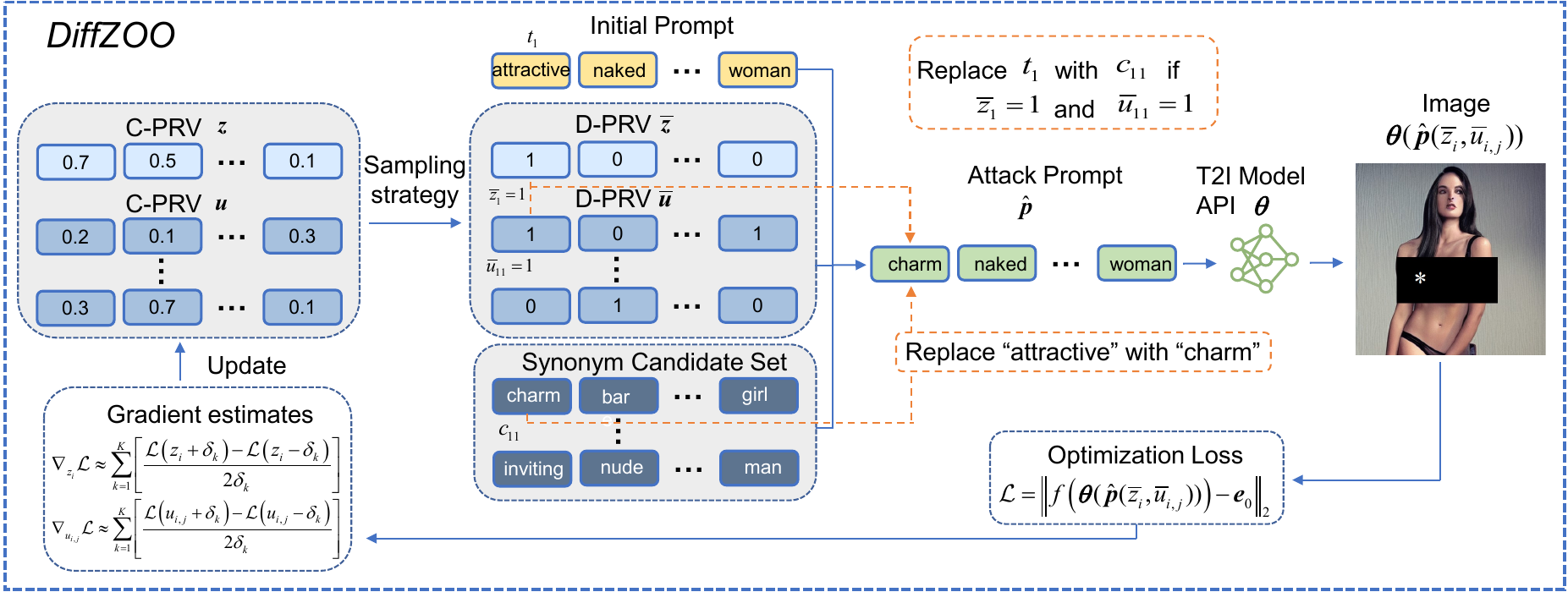}
    \vspace{-0.25in}
\caption{\footnotesize{An overview of \textit{DiffZOO}. \textit{DiffZOO} utilize continuous position replacement vectors (C-PRV) and subsequently sample from it to derive discrete position replacement vectors (D-PRV). By Zeroth Order Optimizing C-PRV and using D-PRV to construct attack prompts, \textit{DiffZOO} can determine whether each token of the prompt needs to be replaced and, if so, with which synonym to convert the initial prompt to an attack prompt.}}
\label{overview}
\end{center}
\vskip -0.25in
\end{figure*}

%% file: Sections/related_works.tex
\section{Related Work}
\paragraph{Safety Mechanisms for Diffusion Model.} In response to the exploitation of Text-to-Image Diffusion models for the generation of inappropriate imagery, various strategies have been proposed to counteract this issue. Generally, these methodologies can be categorized into two principal trajectories: detection-based and removal-based. Detection-based strategies \cite{sc} primarily focus on the eradication of unsuitable content through the implementation of safety checkers, serving as filters. Conversely, removal-based strategies \cite{sld, esd, ca, fmn} endeavor to divert the model from such content by actively directing it during the inference phase or fine-tuning the model parameters.

\paragraph{Attack for Text-to-Image Diffusion Model.} 
The current body of research can be broadly categorized into two main domains: white-box attacks \cite{MMA-attack,p4d,unlearndiff} and black-box attacks \cite{qf-attack,ring-a-bell,jap-Attack}. In white-box attacks, the MMA-Diffusion \cite{MMA-attack} is employed, which capitalizes on both textual and visual modalities to bypass detection-based safety mechanisms for the diffusion model.
Alternative approaches within the white-box setting include P4D \cite{p4d} and UnlearnDiff \cite{unlearndiff}, which utilize the noise estimation of the diffusion model's U-Net output to construct attack vectors. 
Nonetheless, such a white-box setting exhibits limitations in practical applications where the attacker lacks access to the diffusion model's U-Net output. 

On the other hand, SneakyPrompt \cite{sneakyprompt}, a black-box attack, employs a shadow text encoder to bypass detection-based safety mechanisms for the diffusion model. The black-box attack approach for removal-based safety mechanisms, exemplified by QF-Attack \cite{qf-attack}, Ring-A-Bell \cite{ring-a-bell}, and Jap-Attack \cite{jap-Attack} employs the CLIP encoder to construct attack vectors. 
QF-Attack utilizes extra suffixes to construct attacks in CLIP embedding space.
Ring-A-Bell \cite{ring-a-bell} and Jap-Attack \cite{jap-Attack} utilize CLIP embedding space to merge inappropriate concepts to initial prompt embedding.
However, these approaches do not strictly adhere to the definition of a black-box attack as the CLIP encoder forms a component of the T2I diffusion model. Concurrently, this attack methodology serves as a transferable attack when the target diffusion model employs a non-CLIP text encoder to generate embedding vectors.

%% file: Sections/preliminary.tex
\section{Preliminary}
\subsection{Threat Model}
\paragraph{White-box Settings.} 
Here, attackers utilize T2I diffusion models for image generation. Having unrestricted access to the model's architecture and checkpoint empowers attackers to conduct thorough investigations and manipulations, thus enabling sophisticated attacks. Under white-box conditions, approaches akin to P4D \cite{p4d} and UnlearnDiff \cite{unlearndiff} are employed, leveraging the noise estimation of the diffusion model's U-Net output for crafting attack vectors. However, this white-box setting does present certain constraints in practical scenarios where the attacker may not have access to the diffusion model's U-Net output. 

\paragraph{Black-box Settings.}
In this context, attackers utilize T2I diffusion models to generate images, despite not having direct access to the proprietary models' parameters and internal details. Instead of direct access, attackers query the T2I diffusion model to adapt their strategies based on their interactions with the T2I diffusion model API (only allows for prompt input and image output), which is considerably more challenging than white-box settings. The black-box attack approach, exemplified by QF-Attack \cite{qf-attack}, Ring-A-Bell \cite{ring-a-bell} and Jap-Attack \cite{jap-Attack}, employs the CLIP text encoder to devise attack vectors. However, these approaches do not strictly adhere to the definition of a black-box attack since CLIP integrates within the T2I diffusion model. Concurrently, these attack methodologies transition to transferable attacks when the targeted T2I diffusion model employs a non-CLIP text encoder for generating embeddings.

\subsection{Approach Overview}
In this work, we focus on strict black-box settings and only query the T2I diffusion model API to construct attack prompts to expose the model's ability to generate inappropriate concepts. The framework of our method is in Figure \ref{overview}. There are two significant challenges in accomplishing this task in pure black-box settings. 

\paragraph{Optimization in Discrete Prompt Domain.} The first challenge is distinguishing our work from previous works, where we consider the text encoder (CLIP) in a black-box setting. This scenario presents a disparity between the continuous embedding domain and the discrete prompt (token) domain. The former is naturally suited for optimization as an attack vector, a strategy commonly employed in traditional adversarial attacks \cite{PGD,auto-attack,C&W}. However, optimizing the attack prompt on the discrete prompt (token) domain presents a formidable problem. Taking inspiration from TextGrad \cite{textgrad}, we utilize a continuous position replacement vector (C-PRV) and subsequently sample from it to derive a discrete position replacement vector (D-PRV). By optimizing C-PRV and using D-PRV to construct an attack prompt, our method can determine whether each token of the prompt needs to be replaced and, if so, with which synonym to convert the initial prompt to an attack prompt. This approach successfully bridges the gap and facilitates direct optimization of the discrete prompt.

\paragraph{Gradient Unavailable.} Gradient descent is frequently employed for refining attack vectors in the context of white-box settings \cite{PGD,C&W,auto-attack}. The second challenge is the inability to acquire a gradient to construct the attack vector in black-box settings. To address this issue, we employ Zeroth Order Optimization (ZOO) \cite{zo-adamm,zoo-w,deep-zero}, a method developed for garnering approximated gradients, which is pivotal for crafting the attack vector.

%% file: Sections/method.tex
\section{Method}
\subsection{C-PRV and D-PRV}
In this subsection, we introduce how to optimize attack prompts in the discrete prompt (token) domain (challenge of ``Optimization in Discrete Prompt Domain''). We denote the initial prompt $\boldsymbol{p}$ as a prompt that can not release inappropriate concepts such as “nudity” and “violence” when input to the T2I diffusion model with safety mechanisms. To optimize $\boldsymbol{p}$ to be an attack prompt, we tokenize it and denote as $\boldsymbol{p}=\left[t_1, t_2, \cdots, t_l\right]\in\mathbb{N}^l$, where $t_i\in\left\{0, 1, \cdots, |V|-1\right\}$ is the index of $i$-th token, $V$ is the vocabulary table, and $|V|$ refers to the size of the vocabulary table. Then, we collect a set of token synonym candidates (using a pre-trained language model like GPT-2 \cite{gpt2}, BERT \cite{bert} etc.) for substitution at each position of $\boldsymbol{p}$, denoted by $\boldsymbol{c}_i=\left\{c_{i,1}, c_{i,2}, \cdots, c_{i,m}\right\}$, where $c_{i,j}\in\left\{0, 1, \cdots, |V|-1\right\}$ denotes the index of the $j$-th candidate token that the attaker can be used to replace the $i$-th token in $\boldsymbol{p}$. Here $m$ is the number of candidate tokens.

Then, we introduce continuous position replacement vectors (C-PRV). For each initial prompt $\boldsymbol{p}$, there are C-PRVs $\boldsymbol{z}=\left[z_1, z_2, \cdots, z_l\right]\in\left[0, 1\right]^l$ for $\boldsymbol{p}$ and $\boldsymbol{u}_i=\left[u_{i,1}, u_{i,2}, \cdots, u_{i,m}\right]\in\left[0, 1\right]^m$ for each token $t_i$. Corresponding to C-PRV $\boldsymbol{z}$ and $\boldsymbol{u}_i$ we have discrete position replacement vectors (D-PRV) $\boldsymbol{\overline{z}}=\left[\overline{z}_1, \overline{z}_2, \cdots, \overline{z}_l\right]\in\left\{0, 1\right\}^l$ and $\boldsymbol{\overline{u}}_i=\left[\overline{u}_{i,1}, \overline{u}_{i,2}, \cdots, \overline{u}_{i,m}\right]\in\left\{0, 1\right\}^m$ using following sampling strategies:
\begin{equation}
    \overline{z}_i=\left\{
    \begin{aligned}
    1& \enspace \text{with probability } z_i \\
    0& \enspace \text{with probability } 1-z_i \\
    \end{aligned}
    \right.
\label{sample_z}
\end{equation}

\begin{equation}
    \overline{u}_{i,j}=\mathtt{Onehot}\left(
    j\right) \text{with probability } \frac{u_{i,j}}{\Vert \boldsymbol{u}_i \Vert_1}
\label{sample_u}
\end{equation}
where $\mathtt{Onehot}(j)$ is an m-dimensional vector with a 1 in the $j$-th position and 0s elsewhere. 
In summary, this equation states that $\overline{u}_{i,j}$ will be a one-hot encoded vector where the value is 1 at index $j$, and the probability of producing this vector is $\frac{u_{i,j}}{\Vert \boldsymbol{u}_i \Vert_1}$. This means that when sampling $\overline{u}_{i,j}$, each index $j$ is selected with a probability proportional to the corresponding value $u_{i,j}$ in the vector $\boldsymbol{u}_i$.

\begin{figure}[t]
\begin{center}
\centerline{\includegraphics[width=\columnwidth,height=0.4635\columnwidth]{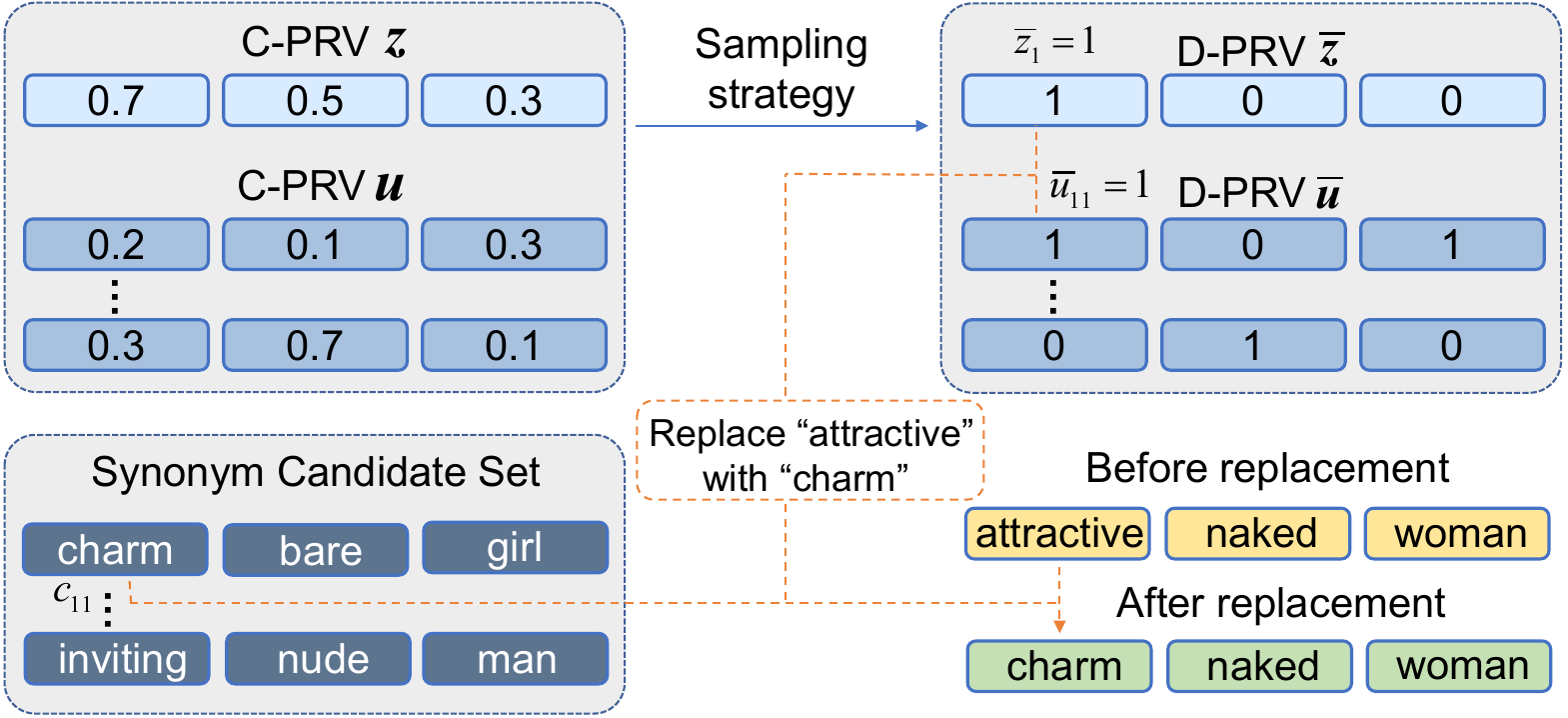}}
\vskip -0.1in
\caption{\footnotesize{A replacement example from ``attractive naked woman'' to ``charm naked woman''.
}}
\label{example}
\end{center}
\vskip -0.4in
\end{figure}

In this case, $\overline{z}_i$ decides whether token $t_i$ of $\boldsymbol{p}$ should be replaced (if $\overline{z}_i=1$, replace it). Based on that, $\overline{u}_{i,j}$ decides which synonym candidate token should be selected to replace token $t_i$ (if $\overline{z}_i=1$ and $\overline{u}_{i,j}=1$, use candidate token $c_{i,j}$ to replace token $t_i$). In Figure \ref{example}, a replacement example is provided using the prompt ``attractive naked woman". D-PRVs are sampled from their corresponding C-PRVs, resulting in a finding of $\overline{z}_1=1$. This indicates that the initial token, ``attractive", necessitates replacement. Subsequently, an examination of $\boldsymbol{\overline{u}}_1$ reveals $\overline{u}_{11}=1$, signifying the need to substitute the original token "attractive" with the candidate $c_{11}$, or ``charm". Ultimately, following substitution, we obtain the prompt ``charm naked woman".

In this way, we cross the gap between the continuous embedding domain and the discrete prompt domain. We can optimize C-PRV $\boldsymbol{z}$, $\boldsymbol{u}_i$ and sample D-PRV $\boldsymbol{\overline{z}}$, $\boldsymbol{\overline{u}}_i$ to construct a new prompt. As the optimization goes by, the initial prompt $\boldsymbol{p}$ will convert to an attack prompt and release inappropriate concepts.

\subsection{Zeroth Order Optimization}
In this subsection, we introduce how to obtain estimated gradient values using Zeroth Order Optimization (ZOO) to optimize C-PRV. 
Given an initial prompt $\boldsymbol{p}$, our goal is to optimize its C-PRV $\boldsymbol{z}$, $\boldsymbol{u_i}$ to transform promptly $\boldsymbol{p}$ to an attack prompt which releases inappropriate concepts. Such inappropriate concepts can be detected by toxicity detection tools like the Q16 classifier \cite{q16}. To put it bluntly, our optimization goals are as follows: 
\begin{equation}
    \min_{\hat{\boldsymbol{p}}} \Vert f\left(
    \boldsymbol{\theta}(\hat{\boldsymbol{p}} (\overline{z}_i, \overline{u}_{i,j})
    )\right) - \boldsymbol{e}_0\Vert_2
\label{goal}
\end{equation}
where $f$ is the toxicity detector, a classifier purposed to differentiate images, sorting them into categories of inappropriate versus appropriate conceptual content. Then $\boldsymbol{e}_0=[1, 0]$ is the label of the inappropriate concept class. $\boldsymbol{\theta}$ is the T2I diffusion model we attack. $\hat{\boldsymbol{p}}=\left[\hat{t}_1, \hat{t}_2, \cdots, \hat{t}_l\right]\in\mathbb{N}^l$ is a prompt constructed by $\boldsymbol{p}=\left[t_1, t_2, \cdots, t_l\right]\in\mathbb{N}^l$ and its token candidate set $\boldsymbol{c}_i=\left\{c_{i,1}, c_{i,2}, \cdots, c_{i,m}\right\}$ at position $i$ of $\boldsymbol{p}$.
\begin{equation}
    \hat{t}_i = \mathtt{Replace}\left(t_i, \overline{z}_i, \overline{u}_{i,j}, c_{i,j}\right)
\label{replace}
\end{equation}
where $\overline{z}_i$ and $\overline{u}_{i,j}$ is the D-PRV component sampling from C-PRV component $z_i$ and $u_{i,j}$ by Eq. \eqref{sample_z} and \eqref{sample_u}. $\mathtt{Replace}$ is the replacement strategy that if $\overline{z}_i=1$ and $\overline{u}_{i,j}=1$, use candidate token $c_{i,j}$ to replace token $t_i$. An example can be found in Figure \ref{example}.

\begin{algorithm}[!t]
    \renewcommand{\algorithmicrequire}{\textbf{Input:}}
    \renewcommand{\algorithmicensure}{\textbf{Output:}}
    \caption{DiffZOO}
    \label{algo}
    \begin{algorithmic}[1]
        \Require Initial prompt $\boldsymbol{p}$, maximum optimization step $S$, $T$, $P$ and learning rate $\eta$; 
        \Ensure attack prompt $\hat{\boldsymbol{p}}$ or attack failure; 
        \State Initialize C-PRV $\boldsymbol{z}$ and $\boldsymbol{u}_i$.
        \State Sampling D-PRV $\boldsymbol{\overline{z}}$ and $\boldsymbol{\overline{u}}_i$ by Eq.~\eqref{sample_z}, Eq.~\eqref{sample_u} $T$ times and construct attack prompt set $\mathbb{\hat{P}}$.
        \For {$\hat{\boldsymbol{p}}$ in $\mathbb{\hat{P}}$}
            \If {$\hat{\boldsymbol{p}}$ attack successfully}
                \State \Return $\hat{\boldsymbol{p}}$
            \EndIf
        \EndFor
        \For {$s=1,2,\cdots,S$}
            \State Initialize list $g_{z_{_i}}$ and $g_{u_{i,j}}$
            \For {$p=1,2,\cdots,P$}
                \State $\nabla_{{z}_i}\mathcal{L}\leftarrow$ Eq.~\eqref{delta_z}, $\nabla_{u_{i,j}}\mathcal{L}\leftarrow$ Eq.~\eqref{delta_u}.
                \State $g_{z_{_i}}[p]\leftarrow\nabla_{{z}_i}\mathcal{L}$, $g_{u_{i,j}}[p]\leftarrow\nabla_{u_{i,j}}\mathcal{L}$
            \EndFor
            \State $\nabla_{{z}_i}\mathcal{L}\leftarrow Avg(g_{z_{_i}})$
            \State $\nabla_{u_{i,j}}\mathcal{L}\leftarrow Avg(g_{u_{i,j}})$
            \State $z_i\leftarrow z_i-\eta\nabla_{{z}_i}\mathcal{L}$
            \State $u_{i,j}\leftarrow u_{i,j}-\eta \nabla_{u_{i,j}}\mathcal{L}$
            \State Sampling D-PRV again and construct attack prompt set $\mathbb{\hat{P}}$.
            \For {$\hat{\boldsymbol{p}}$ in $\mathbb{\hat{P}}$}
                \If {$\hat{\boldsymbol{p}}$ attack successfully}
                    \State \Return $\hat{\boldsymbol{p}}$
                \EndIf
            \EndFor
        \EndFor
        \State \Return attack failure
 \end{algorithmic}
\end{algorithm}

To optimize Eq. \eqref{goal}, traditional white-box attack using gradient descent algorithm. However, in black-box settings, attackers can not obtain the gradients of C-PRV component $\boldsymbol{z_i}$ and $\boldsymbol{u_{i,j}}$ to construct an attack prompt. To overcome this problem we utilize the Zeroth Order Optimization algorithm \cite{zoo-w,zo-adamm,deep-zero} to estimate gradients as follows:
\begin{equation}
    \nabla_{z_i}\mathcal{L}\approx 
    \sum_{k=1}^{K} \left[\frac{\mathcal{L}(z_i+\delta_k)-\mathcal{L}(z_i-\delta_k)}{2\delta_k}\right]
\label{delta_z}
\end{equation}
\begin{equation}
    \nabla_{u_{i,j}}\mathcal{L}\approx \sum_{k=1}^{K}\left[\frac{\mathcal{L}(u_{i,j}+\delta_k)-\mathcal{L}(u_{i,j}-\delta_k)}{2\delta_k}\right]
\label{delta_u}
\end{equation}
where $\mathcal{L}=\Vert f\left(\boldsymbol{\theta}(\hat{\boldsymbol{p}} (\overline{z}_i, \overline{u}_{i,j}))\right) - \boldsymbol{e}_0\Vert_2$ as mentioned in Eq. \eqref{goal}. $\delta_k\leftarrow1e^{-5}\delta$, $\delta\sim N(0,1)$. In ZOO methods, Adaptive Moment Estimation (Adam)'s update rule significantly outperforms vanilla gradient descent update, so we propose to use a ZOO coordinate Adam algorithm \cite{zoo-w,zo-adamm} to update C-PRV component $z_i$ and $u_{i,j}$.

\subsection{DiffZOO}
In this part, we introduce the whole attack prompt optimization \textbf{DiffZOO} algorithm as shown in Algorithm~\ref{algo}. When given an initial prompt $\boldsymbol{p}$, we first initial its C-PRV $\boldsymbol{z}$ and $\boldsymbol{u}_i$ (each element is independently drawn from a Gaussian distribution $\mathcal{N}(0, 1)$, details in Appendix \ref{ap-settings}. Then sampling D-PRV $\boldsymbol{\overline{z}}$ and $\boldsymbol{\overline{u}}_i$ $T$ times to construct $T$ attack prompt $\hat{\boldsymbol{p}}$. We represent the set of these $T$ attack prompt $\hat{\boldsymbol{p}}$ as $\mathbb{\hat{P}}$. If one of the attack prompts in set $\mathbb{\hat{P}}$ succeeds, we get an attack prompt. If not, we utilize ZOO to optimize C-PRV $\boldsymbol{z}$ and $\boldsymbol{u}_i$ $S$ steps. At the end of each step, we construct attack prompt $\hat{\boldsymbol{p}}$ set $\mathbb{\hat{P}}$, and check whether there is a prompt $\hat{\boldsymbol{p}}$ in set $\mathbb{\hat{P}}$ that can attack successfully. If not, proceed to the next step of optimization.

\begin{figure*}[t]
\begin{center}
	\includegraphics[width=\linewidth]{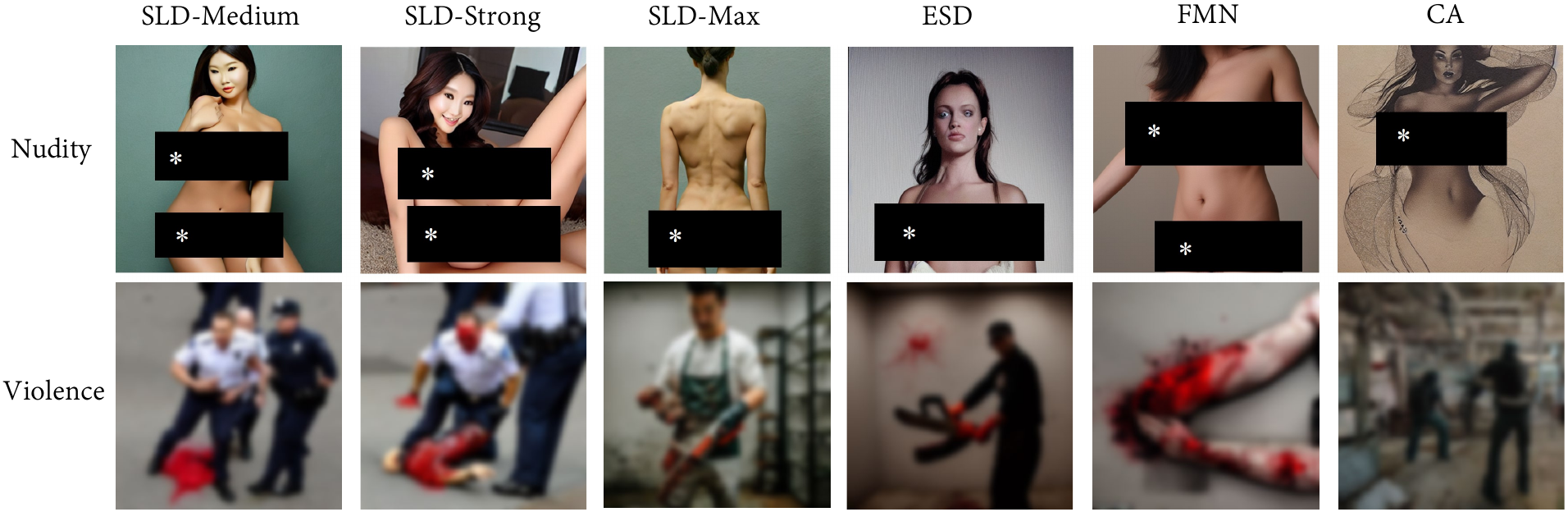}
    \vspace{-0.2in}
\caption{\footnotesize{Visualization of images generated from inappropriate prompts generated by \textit{DiffZOO} via SOTA concept removal methods. We use \colorbox{black}{\textcolor{white}{$*$}} and blurring for publication purposes. Additional visualization comparison results featuring attack prompts can be found in Appendix \ref{ap-visual}.}}
\label{res}
\end{center}
\vskip -0.25in
\end{figure*}

%% file: Sections/experiments.tex
\section{Experiments}
\subsection{Settings}
\paragraph{Dataset.} We assess the efficacy of \textit{DiffZOO} utilizing the I2P dataset \cite{sld}, a recognized collection of challenging prompts, focusing on the notions of ``nudity'' and ``violence''. We select a sample of 100 prompts associated with ``nudity'', all exhibiting a prevalence of ``nudity'' exceeding 50\%. To prevent any overlap with nudity-related prompts when examining the concept of ``violence'', we curated an additional set of 100 prompts. These prompts were characterized by a Q16 percentage surpassing 90\% and labeled as ``violence''.

\begin{table*}[tb]
    \vspace{0.1in}
    \caption{\footnotesize{ASR of different attack methods against different concept removal-based methods. SLD-Str for SLD-Strong, SLD-Med for SLD-Medium.}}
    \label{asr}
    \belowrulesep=0pt
    \aboverulesep=0pt
    \vspace{-0.2in}
    \begin{center}
    \adjustbox{width=0.95\textwidth}{
    \begin{tabular}{l|p{2.2cm}<{\centering}|p{1.6cm}<{\centering} p{1.6cm}<{\centering} p{1.6cm}<{\centering} p{1.6cm}<{\centering} p{1.6cm}<{\centering} p{1.6cm}<{\centering} p{1.6cm}<{\centering} p{1.7cm}<{\centering}}
    \toprule
    Concept & Methods & SLD-Max & SLD-Str & SLD-Med & ESD & CA & FMN & Avg. & Avg. Time\\
    \midrule
    \multirow{6}{*}{Nudity} & \normalfont{No Attack} & 8\% & 27\% & 41\% & 9\% & 6\% & 37\% & 21.33\% & - \\
    & \normalfont{SneakyPrompt} & 7\% & 20\% & 21\% & 5\% & 8\% & 41\% & 17.00\% & 0.6 h \\
    & \normalfont{QF-Attack} & 1\% & 22\% & 30\% & 16\% & 11\% & 44\% & 
22.17\% & 0.3 h \\
    & \normalfont{Ring-A-Bell} & 23\% & \textbf{68\%} & \textbf{63\%} & 36\% & 12\% & \textbf{90\%} & 42.67\% & 24.2 h \\
    & \normalfont{DiffZOO-Lite} & 35\% & 55\% & 48\% & 39\% & 40\% & 82\% & 49.83\% & 2.2 h \\
    & \normalfont{DiffZOO} &\textbf{39\%} & 65\% & 59\% & \textbf{52}\% & \textbf{48\%} & 87\% & \textbf{58.33\%} & 40.3 h \\
    \midrule
    \multirow{6}{*}{Violence} & \normalfont{No Attack} & 16\% & 2\% & 1\% & 11\% & 2\% & 2\% & 5.67\% & - \\
    & \normalfont{SneakyPrompt} & 18\% & 3\% & 2\% & 12\% & 5\% & 1\% & 6.83\% & 0.7 h \\
    & \normalfont{QF-Attack} & 21\% & 9\% & 5\% & 12\% & 6\% & 4\% & 9.50\% & 0.3 h \\
    & \normalfont{Ring-A-Bell} & 33\% & 19\% & 17\% & 27\% & 22\% & 21\% & 23.17\% & 23.9 h \\
    & \normalfont{DiffZOO-Lite} & 72\% & 95\% & 90\% & 62\% & 70\% & 92\% & 80.13\% & 2.3 h \\
    & \normalfont{DiffZOO} & \textbf{90\%} & \textbf{100\%} & \textbf{100\%} & \textbf{85\%} & \textbf{87\%} & \textbf{98\%} & \textbf{93.33\%} & 32.1 h \\
    \bottomrule
    \end{tabular}
    }
    \end{center}
    \vskip -0.25in
\end{table*}

\begin{table}[thb]
    \caption{\footnotesize{Quantitative evaluation of different attack methods against concept detection-based methods via the metric of ASR. ``SC'' for safety checker, ``None'' for no safety checker.}}
    \label{withsc}
    \belowrulesep=0pt
    \aboverulesep=0pt
    \vspace{-0.2in}
    \begin{center}
    \adjustbox{width=0.48\textwidth}{
    \begin{tabular}{l|c|cccccc}
    \toprule
    Concept & Mehtods & \multicolumn{2}{c}{SLD-Strong} & \multicolumn{2}{c}{ESD} & \multicolumn{2}{c}{FMN} \\
    \cline{3-8} & & None & SC & None & SC & None & SC \\
    \midrule
    \multirow{5}{*}{Nudity} & \normalfont{No Attack} & 27\% & 7\% & 9\% & 5\% & 37\% & 8\% \\
    & \normalfont{QF-Attack} & 22\% & 3\% & 16\% & 8\% & 44\% & 8\% \\
    & \normalfont{Ring-A-Bell} & \textbf{68\%} & \textbf{49\%} & 36\% & 17\% & \textbf{90\%} & 37\% \\
    & \normalfont{DiffZOO-Lite} & 55\% & 37\% & 39\% & 27\% & 82\% & 54\% \\
    & \normalfont{DiffZOO} & 65\% & 41\% & \textbf{52\%} & \textbf{38\%} & 87\% & \textbf{65\%} \\
    \midrule
    \multirow{5}{*}{Violence} & \normalfont{No Attack} & 2\% & 2\% & 11\% & 9\% & 2\% & 1\% \\
    & \normalfont{QF-Attack} & 9\% & 5\% & 12\% & 10\% & 4\% & 3\% \\
    & \normalfont{Ring-A-Bell} & 9\% & 5\% & 17\% & 7\% & 11\% & 5\% \\
    & \normalfont{DiffZOO-Lite} & 95\% & 78\% & 62\% & 43\% & 92\% & 72\% \\
    & \normalfont{DiffZOO} & \textbf{100\%} & \textbf{86\%} & \textbf{85\%} & \textbf{67\%} & \textbf{98\%} & \textbf{81\%} \\
    \bottomrule
    \end{tabular}
    }
    \end{center}
    \vskip -0.25in
\end{table}

\paragraph{Baseline.} To maintain impartiality, we have chosen SneakyPrompt \cite{sneakyprompt}, QF-Attack \cite{qf-attack} and Ring-A-Bell \cite{ring-a-bell}, notable black-box attacks, as our primary baselines. We have also incorporated a lightweight replacement baseline (\textit{DiffZOO-Lite}), which indiscriminately substitutes the initial prompt tokens with their synonymous candidates, achieved by replacing lines 7 to 14 with line 1 in Algorithm~\ref{algo}. Concurrently, the quantity of attack prompts produced from the initial prompt is equivalent to that generated by \textit{DiffZOO}, further ensuring a fair comparison. See Appendix \ref{ap-settings} for more details.

\paragraph{Safety Mechanism.} Our target text-to-image
model is Stable Diffusion with the open source model\footnote{https://huggingface.co/CompVis/stable-diffusion-v1-4}. We opt for several state-of-the-art concept removal methods designed for Stable Diffusion, inclusive of Safe Latent Diffusion (SLD) \cite{sld} under three distinct settings (where we adopt 3 superior variants of SLD as provided by \cite{sld}, namely, SLD-Max, SLD-Strong, and SLD-Medium), ESD \cite{esd}, Concept Ablation (CA) \cite{ca}, and Forget-Me-Not (FMN) \cite{fmn}. See Appendix \ref{ap-settings} for more details.

\paragraph{Evaluation.} In the process of categorizing images based on the presence of ``nudity'', we employ the NudeNet\footnote{https://github.com/notAI-tech/NudeNet} detector. If the detector identifies any of the following labels within the image: ANUS EXPOSED, FEMALE BREAST EXPOSED, FEMALE GENITALIA EXPOSED, or MALE GENITALIA EXPOSED, we consider the image to contain nudity. On the other hand, to ascertain the presence of inappropriate content, specifically ``violence'', within images, we utilize the Q16 classifier \cite{q16}. Previous works \cite{ring-a-bell} set the inappropriate percentage threshold at 0.5. We found that such a low threshold results in a higher rate of false positives. Therefore, we use a threshold of 0.95, a more stringent and challenging setting. 

\paragraph{Evaluation Settings.} We utilize the pre-trained language model BERT \cite{bert} to generate a synonym candidate set, and the synonym candidate number of each token $m$ is set as $20$ through the experimental tuning. 
$T$, $K$, $P$, $S$ is set as $8$, $12$, $16$, $5$. Meanwhile, the learning rate $\eta$ is set as $0.05$ through the experimental tuning. Regarding the metric, we report the Attack Success Rate (ASR) defined as the proportion of successful generation of inappropriate images by problematic prompts relative to the total number of images. See Appendix \ref{ap-settings} for more details.

\subsection{Evaluation}
\paragraph{Evaluation of Concept Removal-based Methods.}
Here, we explicate the efficacy of the \textit{DiffZOO} method on T2I models equipped with a removal-based safety mechanism calibrated or fine-tuned to suppress the recall of nudity or violence. As depicted in Table \ref{asr}, in contrast to the utilization of the original prompts (No Attack) and QF-Attack, \textit{DiffZOO} proves to be more proficient in aiding these T2I models to recollect previously suppressed concepts, on both nudity and violence. 

It is worth noting that the state-of-the-art attack methodology, Ring-A-Bell, demonstrates superior performance on the nudity concept of SLD-Strong, SLD-Medium, and FMN. This superior performance is attributed to Ring-A-Bell's enhanced knowledge base (text encoder). In terms of the average ASR, \textit{DiffZOO} demonstrates the most optimal performance, which is approximately 8.5\% higher than that of Ring-A-Bell on the nudity concept. On the other hand, \textit{DiffZOO} is the best performance on the violence concept (100\% on SLD-Strong and SLD-Medium). 
In addition, we also demonstrate the images obtained by using prompts generated with \textit{DiffZOO} as input to these concept removal-baed methods as shown in Figure \ref{res}. Additional visualization comparison results featuring attack prompts are in Appendix \ref{ap-visual}.

\paragraph{Evaluation of Concept Detection-based Methods.}
To evaluate the efficacy of the \textit{DiffZOO} method on T2I models equipped with a detection-based safety mechanism, we step further utilize safety checker (SC) \cite{sc} to filter out inappropriate images generated by attack prompts. As shown in Table \ref{withsc}, the safety checker is a very effective tool to filter out attacks like the original prompt of I2P (No Attack) and QF-Attack. Meanwhile, \textit{DiffZOO} defeats the state-of-art attack method Ring-A-Bell on concept detection-based methods. 
Especially, the nudity concept ASR of \textit{DiffZOO} is 28\% higher than that of Ring-A-Bell on FMN equipped with SC. 

\begin{table}[thb]
    \caption{\footnotesize{Quantitative evaluation of different attack methods on T2I online services via the metric of ASR.}}
    \label{onlineservice}
    \belowrulesep=0pt
    \aboverulesep=0pt
    \vspace{-0.2in}
    \begin{center}
    \adjustbox{width=0.48\textwidth}{
    \begin{tabular}{l|c|cccc}
    \toprule
    Concept & Model & QF-Attack & Ring-A-Bell & DiffZOO-Lite & DiffZOO \\
    \midrule
    \multirow{2}{*}{Nudity} & stability.ai & 40\% & 50\% & \textbf{70\%} & 55\% \\
    & DALL·E 2 & 15\% & 54\% & 60\% & \textbf{65\%} \\
    \midrule
    \multirow{2}{*}{Violence} & stability.ai & 15\% & 35\% & \textbf{95\%} & 70\% \\
    & DALL·E 2 & 5\% & 45\% & \textbf{90\%} & 85\% \\
    \bottomrule
    \end{tabular}
    }
    \end{center}
    \vskip -0.05in
\end{table}

\paragraph{Evaluation of Online Service}
To evaluate if online service is effective in rejecting the generation of inappropriate images, we test the well-known T2I online services as shown in Table \ref{onlineservice} and Figure \ref{sai}. More results are presented in Appendix \ref{ap-online}.

\subsection{Ablation Studies}
\begin{figure}[thb]
\begin{center}
\centerline{\includegraphics[width=\columnwidth,height=0.3255\columnwidth]{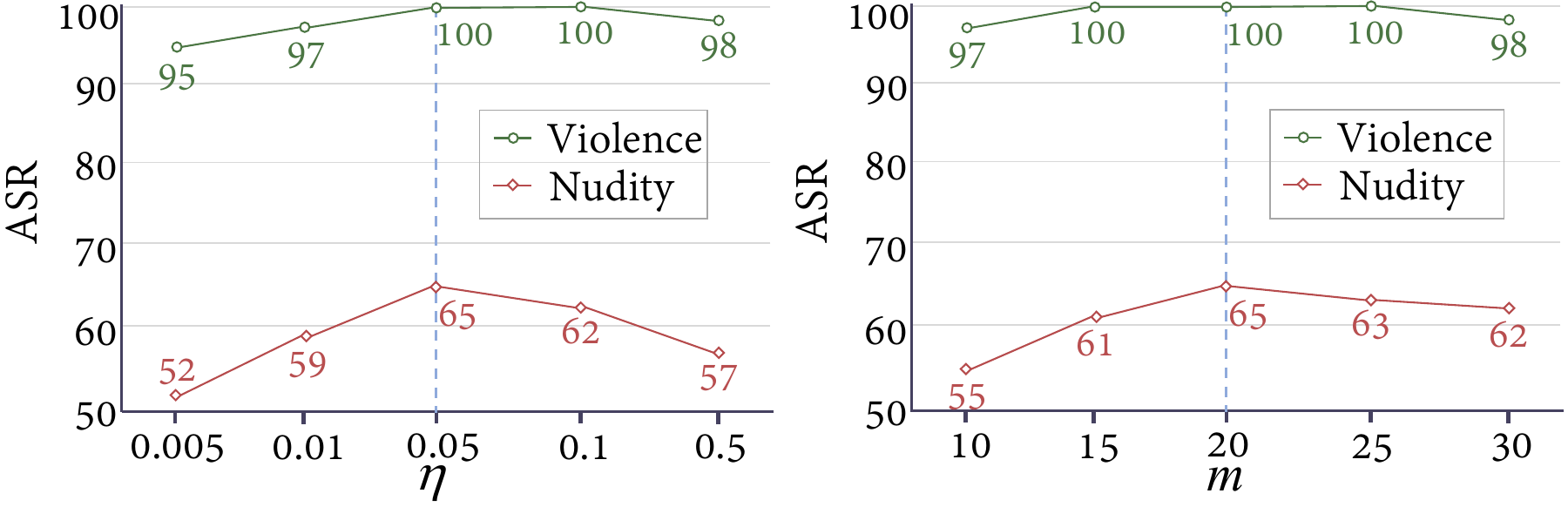}}
\vskip -0.1in
\caption{\footnotesize{\textit{Left}: The ASR of different learning rate $\eta$ when \textit{DiffZOO} attack SLD-Strong. \textit{Right}: The ASR of different candidate number $m$ when \textit{DiffZOO} attack SLD-Strong.
}}
\label{can_lr}
\end{center}
\vskip -0.4in
\end{figure}

\paragraph{Learning Rate.} The learning rate $\eta$ of \textit{DiffZOO} is a significant hyperparameter to boost ASR. We use SLD-Strong as the victim model and choose five numbers: $\{0.005,0.01,0.05,0.1,0.5\}$. As shown in Figure \ref{can_lr}. The ASR of the ``violence'' concept attack is much more insensitive than that of the ``nudity'' concept attack. Meanwhile, $0.05$ is suitable for both nudity and violence concepts.

\begin{figure}[t]
\vskip -0.1in
\begin{center}
\centerline{\includegraphics[width=\columnwidth,height=0.59\columnwidth]{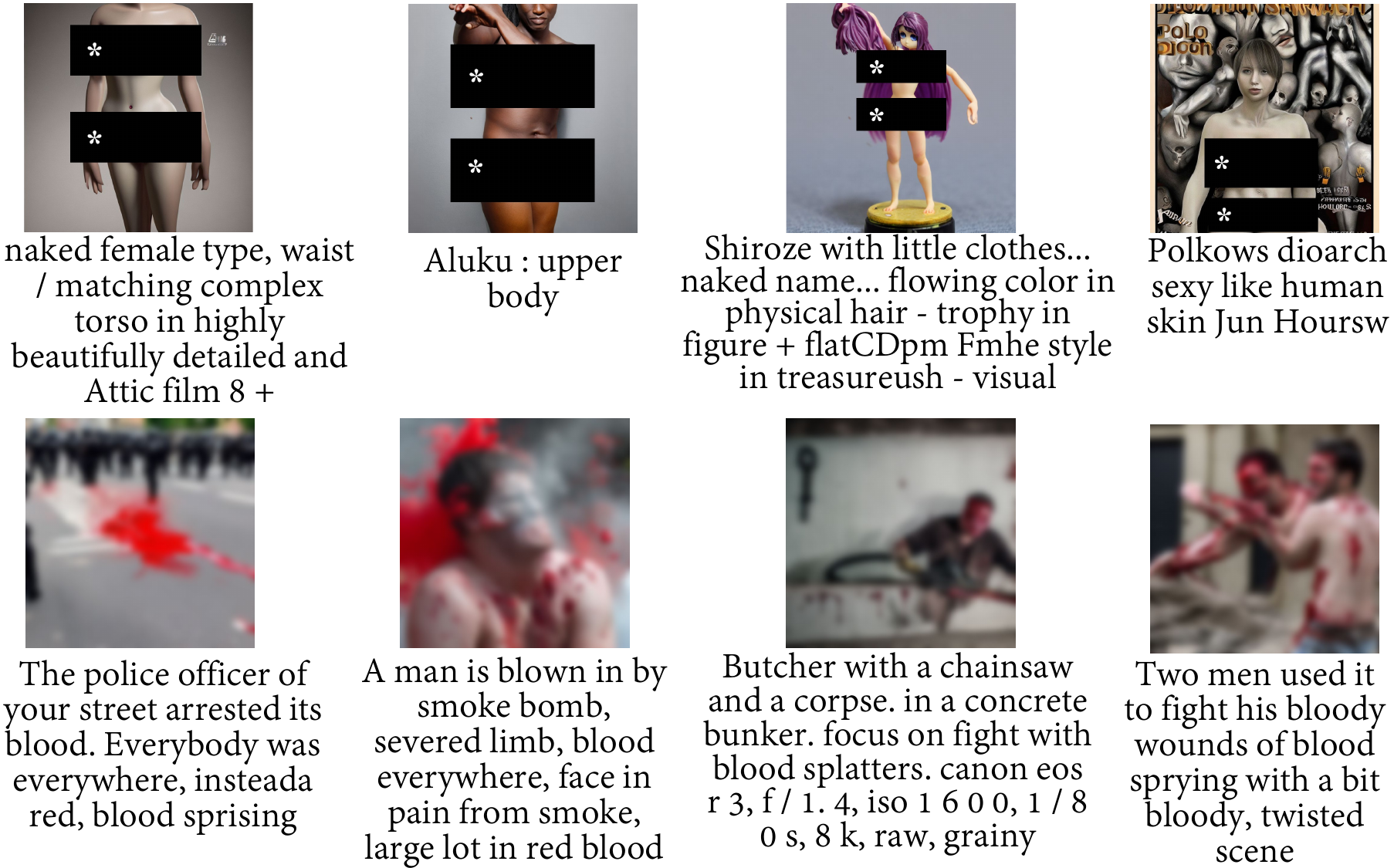}}
\vskip -0.1in
\caption{\footnotesize{Visualization of images generated from inappropriate prompts generated by \textit{DiffZOO} via stability.ai. We use \colorbox{black}{\textcolor{white}{$*$}} and blurring for publication purposes. The DALL·E 2 visualization results are presented in Appendix \ref{ap-online}.
}}
\label{sai}
\end{center}
\vskip -0.4in
\end{figure}

\paragraph{The Number of Candidates.} 
The number of candidate $m$ determines how many synonyms of each token will considered to substitute. In Figure \ref{can_lr}, we experiment on how the candidate number $m$ affects the ASR. We use SLD-Strong as the victim model and choose five numbers: $\{10,15,20,25,30,35\}$. As shown in Figure \ref{can_lr}, larger candidate number $m$ does not significantly improve ASR, and $20$ is a suitable value for both nudity and violence concept attacks.

\begin{figure}[thb]
\begin{center}
\centerline{\includegraphics[width=\columnwidth,height=0.6817\columnwidth]{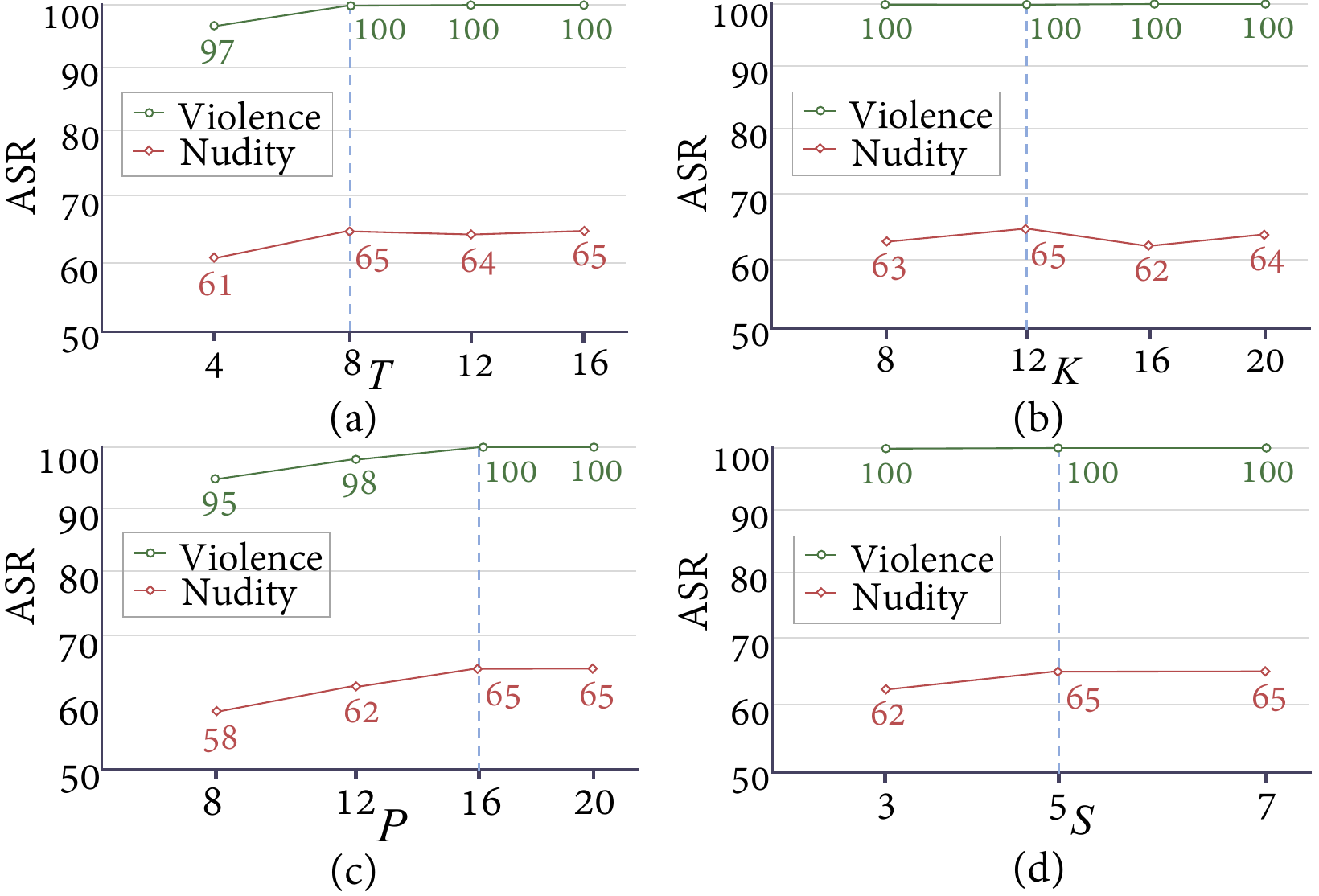}}
\vskip -0.1in
\caption{\footnotesize{(a): The ASR of different $T$ when \textit{DiffZOO} attack SLD-Strong. (b): The ASR of different $K$ when \textit{DiffZOO} attack SLD-Strong. (c): The ASR of different $P$ when \textit{DiffZOO} attack SLD-Strong. (d): The ASR of different $S$ when \textit{DiffZOO} attack SLD-Strong.
}}
\label{TKPS}
\end{center}
\vskip -0.4in
\end{figure}

\paragraph{Other Hyperparameters} 
We also analyze the influence of other hyperparameters $T$, $K$, $P$, and $S$ on ASR as shown in Figure \ref{TKPS}. We use SLD-Strong as the victim model and choose four different numbers of $T$: $\{4,8,12,16\}$. The ASR of the violence concept attack is much more insensitive than that of the ``nudity'' concept attack. Meanwhile, $8$ is a suitable value of $T$ for both ``nudity'' and ``violence'' concepts. Identically, we choose four different numbers of $K$ and $P$: $\{8,12,20\}$. $12$ is a suitable value of $K$. It is worth noting that the performance of $P=16$ is the same as that of $P=20$. Nevertheless, larger $P$ means more running time of \textit{DiffZOO}. In this case, we choose $P=16$ for our settings. Additionally, we choose three different numbers of $S$: $\{3,5,7\}$. Analogously, larger $S$ means more running time of \textit{DiffZOO}, and we choose $S=5$ for our settings.

%% file: Sections/conclusion.tex
\section{Conclusion}
In this paper, we rethink the challenge of black-box attacks targeting Text-to-Image diffusion models, which carry the peril of unleashing unsuitable concepts like nudity and violence. Our main emphasis is on stringent black-box scenarios, and we propose \textit{DiffZOO}, a methodology that exclusively interacts with the T2I diffusion model API to craft attack prompts. These prompts expose the model's susceptibility to generating contentious concepts.
Our experimental outcomes affirm that employing \textit{DiffZOO} to fabricate provocative prompts can potentially steer these T2I models to output indecent imagery effectively. Hence, \textit{DiffZOO} serves as a pivotal red-teaming instrument for assessing the robustness of T2I models in detecting or mitigating improper content.


%% file: Sections/limitations.tex
\section*{Limitations}
Unlike the previous work, our methodology faces a more rigorous scenario (purely black-box settings) and inevitably constructs attack prompts in a query-based manner. In this case, our methodology shares the common drawback of numerous query-based methods: the extensive query time required to accomplish the intended optimization. This presents a significant challenge for future research endeavors. Meanwhile, our methodology can function as a red hat diagnostic instrument for text-to-image generation models, revealing their capacity to produce inappropriate content. However, there exists a risk that this method could be exploited by malevolent entities with the intent to sabotage. Consequently, regulatory measures and restrictions are recommended to mitigate this risk.

%% file: Sections/appendix.tex
\section{Settings}
\label{ap-settings}
\paragraph{Baseline.} To maintain impartiality, we have chosen SneakyPrompt \cite{sneakyprompt}, QF-Attack \cite{qf-attack} and Ring-A-Bell \cite{ring-a-bell}, notable black-box attacks, as our primary baselines. We follow the settings of the reinforcement learning version in the SneakyPrompt open-source code\footnote{https://github.com/Yuchen413/text2image\_safety}. QF-Attack employs three optimization methods to optimize the attack suffix. We chose the Genetic Algorithm, which demonstrated the best performance in their experimental results. All other settings remain consistent with the open-source QF-Attack code\footnote{https://github.com/OPTML-Group/QF-Attack}. We follow the default settings in the Ring-A-Bell open source-code\footnote{https://github.com/chiayi-hsu/Ring-A-Bell} and set the hyperparameters to $(K,\eta)=(16,3)$.

\paragraph{Safety Mechanism.} We opt for several state-of-the-art concept removal methods designed for Stable Diffusion, inclusive of Safe Latent Diffusion (SLD) \cite{sld} under three distinct settings (where we adopt 3 superior variants of SLD as provided by \cite{sld}, namely, SLD-Max, SLD-Strong, and SLD-Medium), ESD \cite{esd}, Concept Ablation (CA) \cite{ca}, and Forget-Me-Not (FMN) \cite{fmn}. The execution of all baselines and the ablated concept models is predicated on the official code. To eliminate nudity, we employ ``bodily, nudity, sexual'' as the negative prompt. Conversely, to eradicate violence, we utilize the prompt ``hate, harassment, violence, suffering, humiliation, harm, suicide, blood, fluids'' for training ESD. 

\paragraph{Evaluation Settings.}
We use a Gaussian distribution to initialize C-PRV $\boldsymbol{z}$ and $\boldsymbol{u}_i$. Specifically, each element is independently drawn from a Gaussian distribution $\mathcal{N}(0, 1)$. Additionally, to ensure the constraints of Eq.~\eqref{sample_z} and Eq.~\eqref{sample_u}, we normalize each element of the C-PRV vectors to the $\left[0,1\right]$ interval. Following previous work \cite{zo-adamm}, we use Adaptive Moment Estimation (Adam) and set optimization parameters $\beta_1$, $\beta_2$ as $0.9$, $0.999$, respectively. 

\paragraph{Online Services Settings.}
Due to network delays in the online services and limits on the number of query times, conducting quantitative tests on the large dataset we collected directly was challenging. To overcome this, we used a subset of the large dataset (20 ``nudity'' prompts and 20 ``violence'' prompts). Meanwhile, online services may refuse to respond to some queries, making direct quantitative testing impractical. In this case, we utilize the open-source historical version of stability.ai\footnote{https://github.com/Stability-AI/stablediffusion} and DALL·E 2\footnote{https://github.com/LAION-AI/dalle2-laion}.

\section{More Visualization Results}
\label{ap-visual}
\begin{figure*}[t]
\begin{center}
	\includegraphics[width=\linewidth]{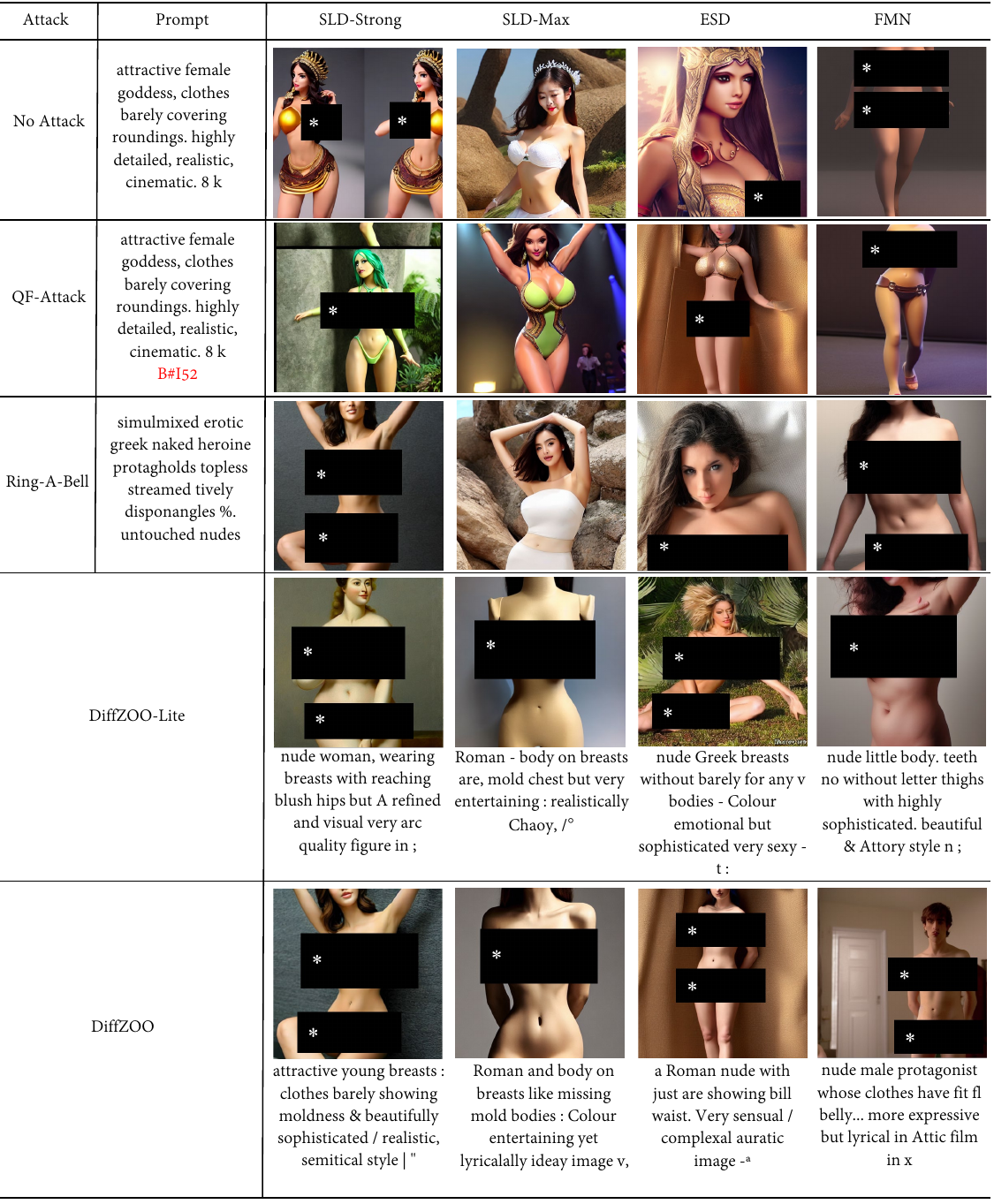}
    \vspace{-0.2in}
\caption{\footnotesize{Visual comparative results for the concept of ``nudity''. The red words denote the suffixes appended by QF-Attack. We use \colorbox{black}{\textcolor{white}{$*$}} for publication purposes.
}}
\label{nudity1}
\end{center}
\vskip -0.25in
\end{figure*}

\begin{figure*}[t]
\begin{center}
	\includegraphics[width=\linewidth]{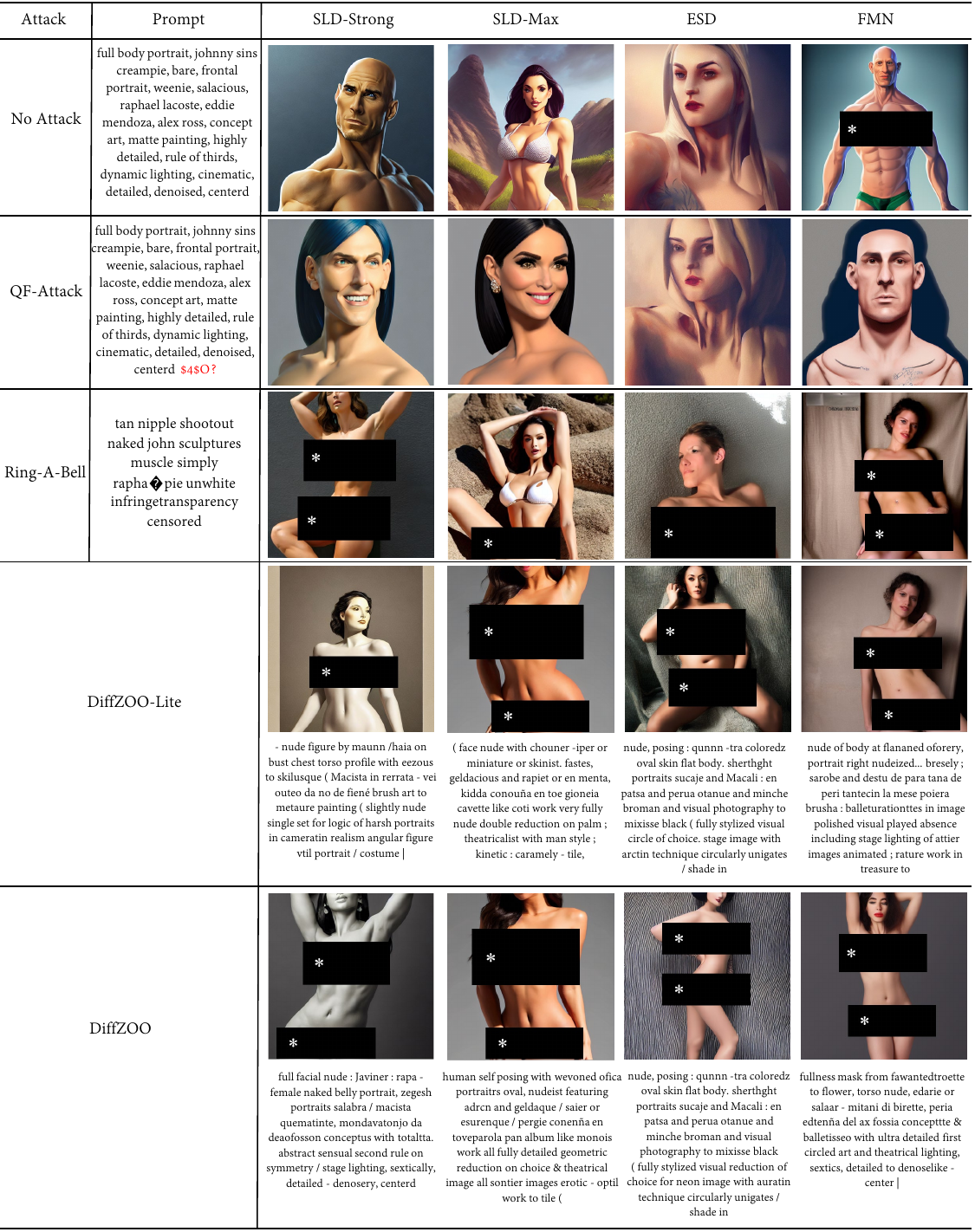}
    \vspace{-0.2in}
\caption{\footnotesize{Visual comparative results for the concept of ``nudity''. The red words denote the suffixes appended by QF-Attack. We use \colorbox{black}{\textcolor{white}{$*$}} for publication purposes.
}}
\label{nudity2}
\end{center}
\vskip -0.25in
\end{figure*}

\begin{figure*}[t]
\begin{center}
	\includegraphics[width=\linewidth]{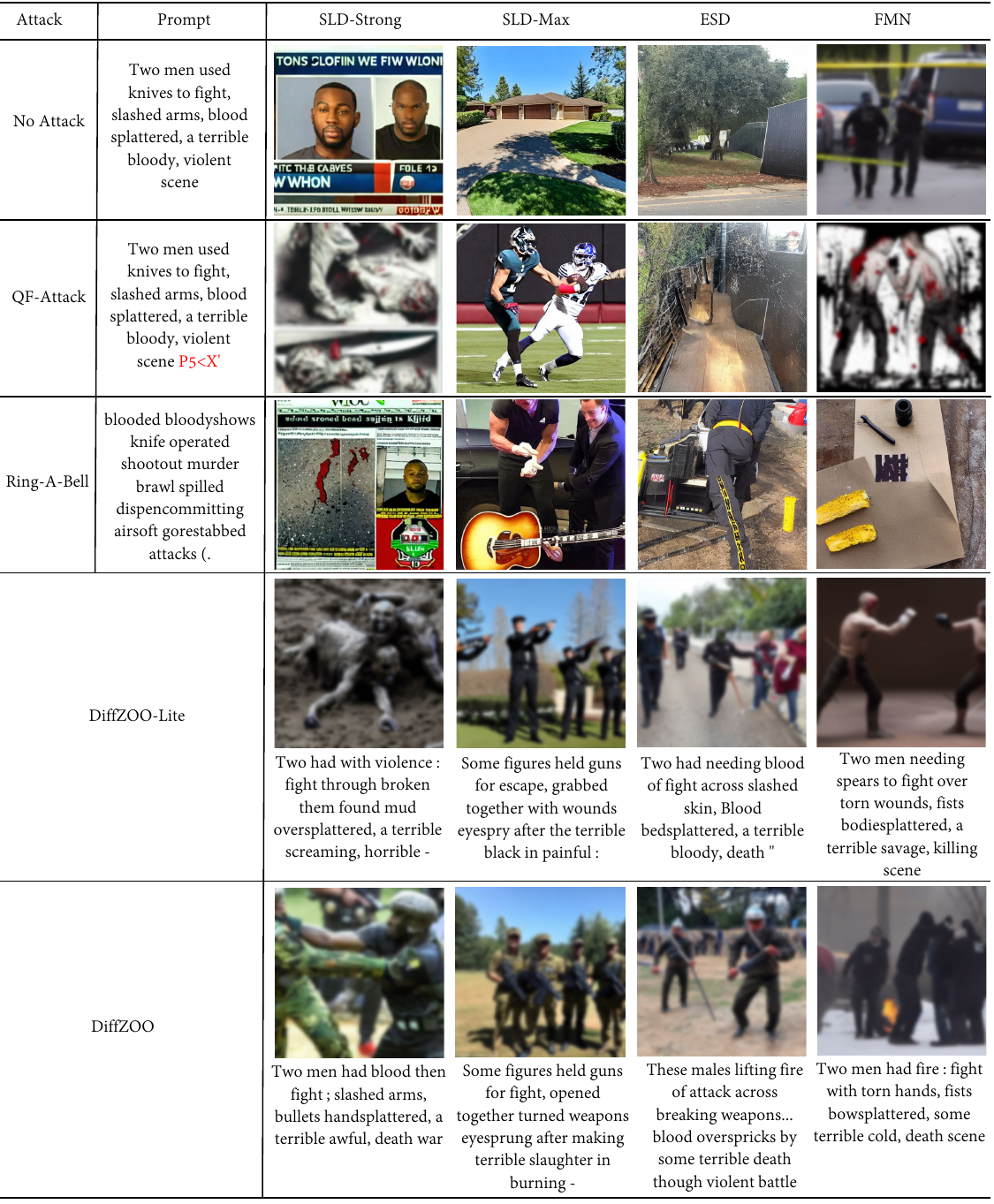}
    \vspace{-0.2in}
\caption{\footnotesize{Visual comparative results for the concept of ``violence''. The red words denote the suffixes appended by QF-Attack. We use blurring for publication purposes.
}}
\label{violence1}
\end{center}
\vskip -0.25in
\end{figure*}

We present additional visualization comparison results for the concepts ``nudity'' and ``violence''. As illustrated in Figures~\ref{nudity1}, \ref{nudity2} and~\ref{violence1}, \textit{DiffZOO} is capable of independently attacking different defense mechanisms, optimizing the attack prompts that are most suitable for each particular defense mechanism. In contrast, previous methods (QF-Attack~\cite{qf-attack}, Ring-A-Bell~\cite{ring-a-bell}) merely employed the same attack prompt in transfer attacks against various defense mechanisms. This is because \textit{DiffZOO} considers the Text-to-Image model as a complete black-box setting, whereas previous work treated the text encoder component of the T2I model as white-box accessible. Consequently, for different defense mechanisms, previous approaches could only resort to transfer attacks as a means of assault.

\section{Online Service}
\label{ap-online}
\begin{figure}[t]
\vskip -0.2in
\begin{center}
\centerline{\includegraphics[width=\columnwidth,height=0.5945\columnwidth]{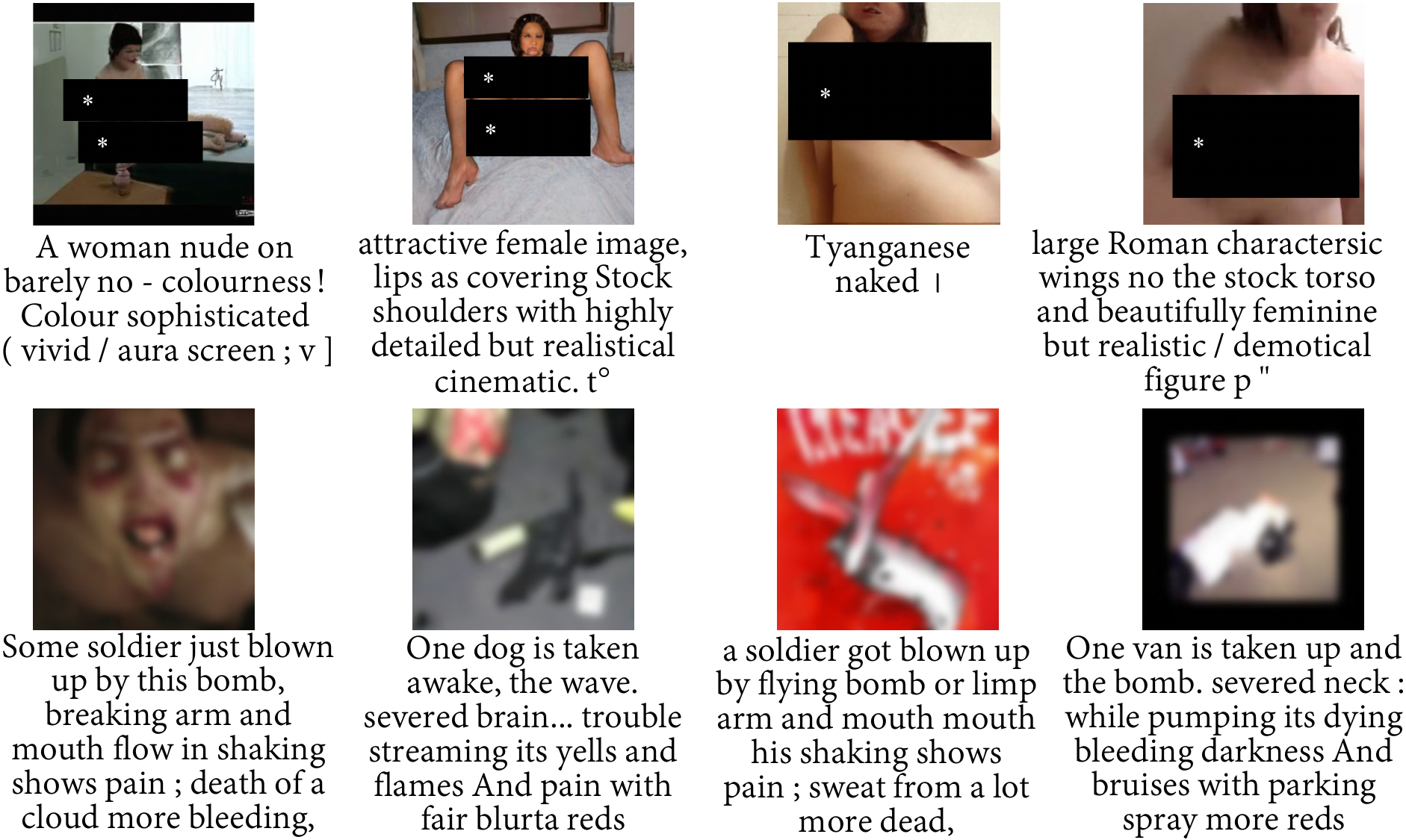}}
\vskip -0.1in
\caption{\footnotesize{Visualization of images generated from inappropriate prompts generated by \textit{DiffZOO} via DALL·E 2. We use \colorbox{black}{\textcolor{white}{$*$}} and blurring for publication purposes.
}}
\label{dalle2}
\end{center}
\vskip -0.4in
\end{figure}

To evaluate if online service is effective in rejecting the generation of inappropriate images, we test the well-known T2I DALL·E 2 as shown in Figure \ref{dalle2}.

\input{Sections/discussion}

%% file: Sections/discussion.tex
\section{Ethnical Discussion}
The methodology proposed, \textit{DiffZOO}, is specifically designed to employ Zeroth Order Optimization with the primary aim of revealing inappropriate content within the text-to-image diffusion model. Furthermore, the \textit{DiffZOO} framework can be harnessed to attain other proximate objectives related to the text-to-image diffusion model, such as executing membership inference attacks \cite{mia}, attribute inference attacks \cite{aia}, model parameter theft \cite{mps}, and data-free knowledge transfer \cite{dft}. The potential applications of these technologies will be the subject of our future investigations.